# Bayesian Weapon System Reliability Modeling with Cox-Weibull Neural Network

Benny Cheng[*], PhD, Naval Surface Warfare Center- Corona
Michael Potter[*], MS, Naval Surface Warfare Center – Corona
* Denotes equal contributions

## ABSTRACT

We propose to integrate weapon system features (such as weapon system manufacturer, deployment time and location, storage time and location, etc.) into a parameterized Cox-Weibull [1] reliability model via a neural network, like DeepSurv [2], to improve predictive maintenance. In parallel, we develop an alternative Bayesian model by parameterizing the Weibull parameters with a neural network and employing dropout methods such as Monte-Carlo (MC)-dropout for comparative purposes. Due to data collection procedures in weapon system testing we employ a novel interval-censored log-likelihood which incorporates Monte-Carlo Markov Chain (MCMC) [3] sampling of the Weibull parameters during gradient descent optimization. We compare classification metrics such as receiver operator curve (ROC) area under the curve (AUC), precision-recall (PR) AUC, and F scores to show our model generally outperforms traditional powerful models such as XGBoost and the current standard conditional Weibull probability density estimation model.

## 1 INTRODUCTION

Survival Analysis techniques are ubiquitous in medical applications for survival curve estimation from clinical trials, treatment recommendations to extend life expectancy, and covariate exploration, to name a few [1,4,5,6]. Recent trends in Machine Learning/Artificial Intelligence (ML/AI) have transitioned from standard survival models such as the linear Cox proportional hazards model to deep nonlinear Cox proportional hazards models to capture the increasingly complex relationships between covariates [2,7, 8].

In the case of weapon systems, current failure time reliability modeling practices utilize only a few features, such as the weapon age and the last time since recertification test (tslrt). As a case study of incorporating more informative features, we assume a nonhomogeneous Poisson process (NHPP) and model the reliability of a weapon system as a Bayesian Weibull probability model with said features, where the posterior of the Weibull parameters is estimated with Monte-Carlo Markov Chain (MCMC) sampling with carefully chosen prior distributions on the parameters.

Understanding population failure statistics via Weibull model parameters is important, but this formulation alone can lead to less-than-optimal individual predictive maintenance because many likely important features are not incorporated, such as weapon manufacturers, times at location, storage locations, and so on. To address these issues, we use our Weibull inductive bias (including the deployed population failure statistics) and incorporate informative weapon systems features via a Cox-Weibull survival analysis that uses a neural network to input the complex weapon system features. To the best of our knowledge, we are the first to develop Bayesian Deep Cox-Weibull reliability models for weapon systems in the Navy.

The paper is organized as follows: section 3 discusses the data formulation and notation for our weapon system dataset. Section 4 describes our novel model architecture and the comparative baseline models. Section 5 outlines how we formulate a Bayesian model with the Neural Network aspect, and section 6 outlines how we train the models. Section 7 and section 8 highlight our results compared to benchmark Survival Analysis models, and section 9 outlines future work.

## 2 RELATED WORK

The current predominant practice in failure time reliability modeling with multiple features involve linear regressions of the covariates as parameters of the reliability model [9,10]. For example, a popular method with Weibull (scale, shape) parameters $(\lambda, c)$ is to incorporate the features $x$ as $\lambda = \lambda(x) = e^{x^T \beta}$, where $\beta$ are additional parameters to be estimated. Similar techniques are applied for other common models such as logistic and Poisson reliability models [11]. Proportional hazard (PH) models for reliability are less common and found mostly in medical applications to survival data [12]. All these models are limited by the number of features that they can handle, including interaction terms, and with only a handful of significant covariates at most. Furthermore, for Bayesian inference, assigning priors to all the feature parameters is quite impractical. With hundreds of features that are available for analysis, it is natural to approach this task from an ML/AI perspective.

## 3 NOTATION AND DATA FORMULATION

- **X**: matrix
- **x:** vector
- x: scalar
- $diag([1,2,\ldots,N]) = \begin{bmatrix} 1 & 0 & 0 \\ 0 & \ddots & 0 \\ 0 & 0 & N \end{bmatrix}$
- { }: set
- $x_i$: individual record of data $\in \mathbb{R}^{1 \times d}$
- $T$: random variable denoting weapon system time to failure

The weapon system dataset contains *n* samples and follows a relational form $= (X \in \mathbb{R}^{n \times d}, y \in \{0,1\}^{n \times 1}, t_1 \in \mathbb{R}^{n \times 1}, t_2 \in \mathbb{R}^{n \times 1})$, where $X$ denotes the matrix of *d*-dimensional features of all samples. Each binary scalar element of labels $y$ denotes a weapon system functionality test $y_i \in \{0,1\}$, where 1 denotes failure and 0 denotes pass. The weapon system total age is denoted by $t_{2i} \in [0, \infty]$, and the time interval the weapon system is in the fleet between recertification tests (the outbound test and the inbound test) is denoted by *tslrt*. The age of the weapon system at the previous recertification test is then $t_1 = t_2 - tslrt$, where $t_{1i} \in [0, \infty]$. An example timeline of the weapon system data collection is shown in Figure 1, where $t_1 = t_{current\ outbound}$ and $t_2 = t_{current\ inbound}$. Due to nature of the data collection process, our data is interval censored (and right censored) because we do not know when the weapon system failed between the current outbound and current inbound test. A dataset sample $(x_i, y_i)$ denotes the measured features and test result of a specific weapon system at time $t_{2i}$. We have over 250 continuous, ordinal, and categorical features as inputs for n>10000 samples. More details of the weapon system may not be disclosed due to the sensitive nature of Navy systems.

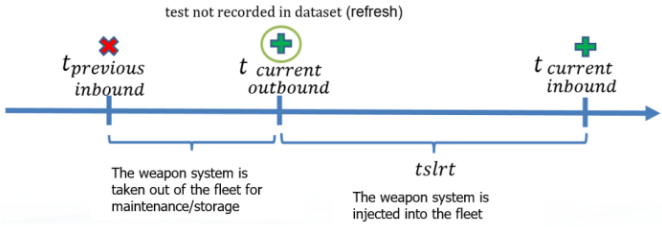

*Figure 1. Weapon System Lifecycle, where the green plus indicates a passed test, and a red x denotes a failure*

When training the neural network to learn the Cox-Weibull reliability model, the data is batched for gradient descent, where a batch of data is denoted as $X_b = \{(x_i, y_i, t_{1i}, t_{2i}): i \sim D\ for\ i = 1:b\}$ where $i$ is randomly sampled with replacement from the dataset until $|X_b| =$ batch size $b$.

## 4 SURVIVAL ANALYSIS FOR COX-WEIBULL MODEL

We formulate the initial probability that the weapon system will pass the test up to time $t_2$ via the Weibull probability density function (pdf) in Equation (1):

$$S_0(t_2) = p[T > t_2] = e^{-(\lambda t_2)^c}, \quad (1)$$

where $\lambda, c$ are the rate and shape parameters of the Weibull pdf, respectively.

We incorporate weapon system features at time $t_1$, assuming they will remain constant in the interval $[t_1, t_2]$, such as time at sea, manufacturer, time inbound, to formulate a non-linear Cox-Weibull survival model [13,2] given by Equation (2):

$$p[T > t_2|x] = S_0(t_2)^{e^{H_\theta(x)}} = e^{-(\lambda t_2)^c e^{H_\theta(x)}}, \quad (2)$$

where $H_\theta$ is a shallow fully-connected Neural Network, and we denote $score = H_\theta(x)$.

We formulate a conditional Cox-Weibull survival model (Equation (3)) to ensure that we only consider the time related to the weapon system circulating in the fleet, and not in storage/maintenance, as a weapon system is fixed and retested for certification until pass at $t_1$.

$$p[T > t_2|T > t_1, x] = e^{(t_1^c - t_2^c)\lambda^c e^{H_\theta(x)}} \quad (3)$$

To classify individual weapon systems as pass or fail at time $t_2$, we threshold the conditional survival probability as:

$$\hat{y} = p[T > t_2|T > t_1, x] > .5 \quad (4)$$

Following the Anderson-Gill recurrent event method, we assume no specific dependence structure among the recurrent intra-weapon system events, and therefore, our model assumes that each unique weapon system has a risk for a $j^{th}$ failure from $t = 0$ onwards irrespective of whether a $(j-1)^{th}$ event has already been observed [14,15,16]. Thus, for each unique weapon system, we assume that the instantaneous risk to experience an event at time $t$ remains the same irrespective of whether previous events have occurred or not, which is a common technique in conditional or marginal survival models for machinery. Therefore, we say recurrent events are assumed to be conditionally independent given the features at time $t$, motivated by the fact that at recertification time, weapon systems are fixed/adjusted to act as if the event (failure/issues) has never occurred and the history is encompassed in the features; this is also a Markovian assumption.

## 5 NEURAL NETWORK ARCHITECTURES

We have two formulations for the Neural Network: 1. the Weibull parameters $\lambda$, c are estimated via MCMC posterior sampling (Figure 2) which we denote as MCMC Cox-Weibull, and 2. the Weibull parameters are estimated by another Neural Network (Figure 3) which we denote as Monte-Carlo(MC)-Dropout Cox-Weibull.

*MCMC Posterior Sampling:* We use the No-U-Turn Sampler [17] for MCMC sampling of the posterior on the Weibull parameters $\lambda, c$ (Equation (5)) with a burn-in of 2000 samples and default number of draws (500). We observe a low Monte Carlo Standard Error (MCSE) and a posterior predictive p-value close to 0.5. The MCMC posterior samples are generated prior to gradient descent optimization of the neural network, where the number of samples needed is $N =$ # $Batches \times$ # $Epochs$.

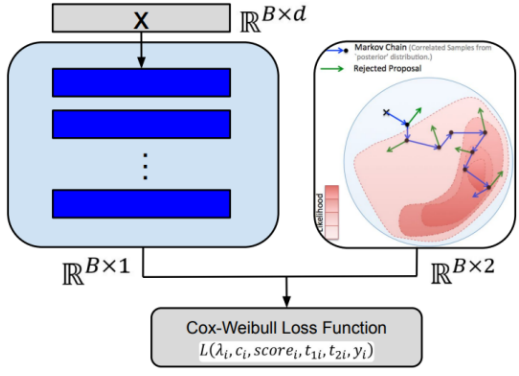

*Figure 2. MCMC Cox-Weibull Model Architecture*

The posterior on the Weibull parameters is given by:

$$p[\lambda, c | t_1, t_2, y] \sim p[y | t_1, t_2, \lambda, c] p[c] p[\lambda|c] \quad (5)$$

where the observed data is the training data.

Due to the proprietary nature of Navy weapon systems, we cannot disclose the prior distributions. The *score* is calculated from a shallow Neural Network $H_\theta$.

*MC-Dropout:* Rather than using MCMC sampling to acquire the Weibull parameters, we develop a multi-task Neural Network (Figure 3) that simultaneously predicts the Weibull parameters $\lambda, c$ and the *score*. MC-Dropout Cox-Weibull has a base sequence of fully connected layers $H_\Theta$, a task branch $H_\phi$ that estimates the Weibull parameters, and a *score* branch $H_\theta$. The base branch outputs a latent feature representation, $z = H_\Theta(x)$, that is input to each task branch.

Before to every fully connected layer in the Neural Network, we add dropout ($p_l$), where $p_l$ denotes the dropout rate for layer $l$ in the Neural Network. Since we assume there are true underlying Weibull parameters $\lambda, c$ that generates the pass/failure data as a function of time, we average the $\lambda_i, c_i$ generated for each weapon system sample $x_i$ in the batch to produce population wide statistics.

$$H_\phi(Z_b) = \lambda_b, c_b \, ; \, \lambda_b, c_b \in \mathbb{R}^{b \times 1} \quad (6)$$
$$[\lambda, c] = [\overline{\lambda_b}, \overline{c_b}] \in \mathbb{R}^{1 \times 2} \quad (7)$$

Unlike the traditional Cox-partial log likelihood objective function (Cox Partial Likelihood) used to optimize Cox-Weibull survival models, which conveniently drops the Weibull parameters from the optimization formulation, MC-Dropout Cox-Weibull optimization includes the parametrized Weibull parameters.

*Dropout:* [18] has shown that by adding dropout before every fully-connected layer in a Neural Network that the Neural Network objective function may approximate a Bayesian Gaussian Process and provide credible intervals for each prediction. Let us denote $\Phi = \{\phi, \theta, \Theta\}$ as all the neural network parameters. Then,

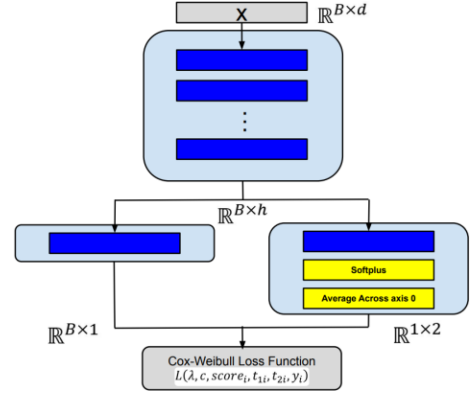

*Figure 3. MC-Dropout Cox-Weibull Model Architecture*

$$M \sim diag\left(\{k_{lj} \sim Bernoulli(p_l)\}_{j=1}^{\# \, of \, neurons \, in \, layer \, l}\right) \quad (8)$$
$$\Phi'_l = \Phi_l M \quad (9)$$

Furthermore, we assume a small weight decay is employed (Equation (10)), where $\alpha$ controls the strength of the weight decay.

$$R(\Phi_l) = \alpha \|\Phi_l\|_2^2 \quad (10)$$

A higher alpha corresponds to a stronger regularization on the Neural Network weights.

Then, [18] shows that weight decay and dropout objective function is equivalent to minimizing the Kullback-Leibler divergence between the variational approximation of the posterior of a deep Gaussian process. Therefore, adding dropout gives a variational approximation on the intractable posterior on the Neural Network weight distribution. Despite the approximating distribution being a sum of two-point masses, where each point mass has zero variance, the mixture does not have zero variance. Thus, the variance of the Bernoulli random variable being transformed through the Neural Network enables complex distributions over the score, $H_{\Phi'}(x)$ [18].

## 6 BAYESIAN INFERENCE

As suggested by the model architecture names, we have two methods for developing Bayesian models: MCMC posterior sampling on the Weibull parameters and the employing dropout before every fully connected layer in the neural network (Figure 2) or applying dropout before every fully connected layer to estimate $\lambda, k$ and score (Figure 3).

During inference time, we employ Bayesian Model Averaging (BMA) with respect to the Weibull parameters and/or the Neural Network weights ($\Phi = \{\theta, \Theta\}$ from here on) for more robust predictions (Equations (11)-(12)).

$$p_{BMA}[y = 1 | x_{new}, t_{1 \, new}, t_{2 \, new}, y] =$$
$$\int p[y = 1 | \Phi, \lambda, c, x_{new}, t_{1 \, new}, t_{2 \, new}, y] q[\Phi] p[\lambda, c | t_1, t_2, y] \, d\Phi d\lambda \, dc \quad (11)$$

$$\approx \frac{1}{U} \sum_{i=1}^{U} p[y = 1 | \Phi'_i, \lambda_i, c_i, x_{new}, t_{1 \, new}, t_{2 \, new}, y]$$

$$= \frac{1}{U}\sum_{i=1}^{U} 1 - e^{\left(t_{1\ new}^{c_i}-t_{2\ new}^{c_i}\right)\lambda_i^{c_i} e^{H_{\Phi_i'}(x_{new})}} \quad (12)$$

where the observed data in the posterior distributions is the training data.

## 7 TRAINING

We partition the training and test data into an 80%-20% split respectively, and then use 20% of the training data as the validation data for early stopping during stochastic gradient descent (SGD). We stratify the data partitions by the binned weapon system age and the recertification test results. We keep a running window average of 5 epochs on the validation ROC-AUC and stop at the highest average ROC-AUC. We use a large batch size of 512 to increase the likelihood of sampling negative samples in each batch due to the highly unbalanced distribution of labels in the dataset, and the ADAM optimizer for the log likelihood optimization. The MCMC + MC-Dropout Cox-Weibull architecture consists of three hidden layers (50-25-25 neurons) with Leaky ReLU activations except at the last hidden layer, and a dropout rate of 0.01. MC-Dropout Cox-Weibull's has a base branch of two hidden layers (50-25 neurons) with Leaky ReLU activations, a task branch of one hidden layer (25 neurons) with a softplus activation ($\lambda \geq 0, c \geq 0$), and a *score* branch with a hidden layer (25 neuron), and a dropout rate of 0.1.

The log-likelihood objective function (we write for the combination of MCMC + Dropout) is cross-entropy (CE) with $P(y=1)$ from Equation (3) and $\ell_2$ regularization on the Neural Network weights $\Phi$:

$$\min_{\Phi} -(\sum_i BCE\left(p_{BMA}[y=1|x_i, t_{1\ i}, t_{2\ i}], y_i\right) - \alpha\|\Phi\|_2^2) \quad (13)$$

where BCE is the binary cross entropy.

Depending on the model architecture selected, small modifications on Equation (13) are made such as how $\lambda, c$ are calculated and the set of Neural Network weights $\Phi$ containing a combination of $\{\phi, \theta, \Theta\}$.

## 8 RESULTS

All the results in Table 1 are calculated as the mean percent change with respect to the conditional Weibull pdf ($V_1$) to $V_2$:

$$\%\Delta = \frac{(V_2 - V_1)}{|V_1|} \times 100 \quad (14)$$

All results were averaged over 10 runs, with each run having a different data partition (and random number seed). Compared to the current practice of conditional Weibull pdf, we show that MCMC posterior sampling and MC-Dropout parameterizing a Cox-Weibull survival model ($\lambda, c$ are sampled from MCMC, not $H_\phi$) gives superior performance with respect to all metrics. Moreover, our approach leads to large gains in all positive class metrics, along with equal performance in all negative class metrics when comparing to the state-of-the-art XGBoost classifier and stochastic variational inference mean field approximation (SVI MFA). We follow [1] to formulate the Weibull regression model for survival analysis as a comparison baseline and minimize the ELBO using stochastic gradient descent. As the dataset contains an order of magnitude fewer fail test results compared to pass test results, and therefore, having a higher precision, recall, and F score for the positive class is critical. We have slightly lower ROC-AUC and Precision-Recall (PR)-AUC compared to XGBoost. That said, when subject matter experts (SMEs) select a subset of important features (approximately 20 features from the 250 features), the ROC-AUC and PR-AUC become higher than XGBoost (Table 2). As expected, all methods outperform the conditional Weibull pdf, because the conditional Weibull pdf only uses weapon system age and tslrt as features.

We generate survival curves (Figure 4), and show that introducing features induces a similar, but slightly lower, reliability over time than the conditioned Weibull probability density without features, which is expected according to SMEs.

*Table 1 - Relative Percent Change with Respect to the conditional Weibull pdf. Bold is best, underline is second best.*

| Model | P1 | R1 | F1 | P0 | R0 | F0 | ROC AUC | PR AUC | $C^{td}$ |
|---|---|---|---|---|---|---|---|---|---|
| Conditional Weibull pdf | 0 | <u>0</u> | <u>0</u> | 0 | <u>0</u> | <u>0</u> | 0 | 0 | 0 |
| MCMC | <u>369.44</u> | **16.67** | **16.67** | 0 | 1.01 | 0.62 | 9.83 | 42.43 | <u>5.69</u> |
| MC-Dropout | -100 | -100 | -100 | 0 | 1.01 | 0.62 | 5.61 | 16.30 | -2.10 |
| **MCMC + MC-Dropout** | 344.44 | **16.67** | **16.67** | 0 | 1.01 | 0.62 | <u>9.84</u> | <u>43.82</u> | **5.75** |
| XGBoost | -100 | -100 | -100 | 0 | 1.01 | 0.62 | **12.8** | **52.05** | 2.19 |
| SVI MFA | **547.22** | -33.33 | -33.33 | 0 | 1.01 | 0.62 | 6.13 | 19.65 | 0.19 |

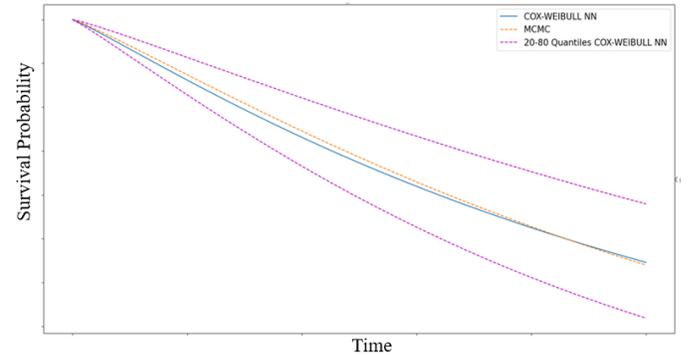

*Figure 2. Survival Curves for conditional Weibull pdf and MCMC + MC-Dropout Cox-Weibull Model on the subset of selected features*

*Table 2 - ROC-AUC and PR-AUC for subset of selected features for Relative Percent Change with Respect to the conditional Weibull pdf.*

*Bold is best, underline is second best.*

| Model | ROC AUC | PR AUC |
|---|---|---|
| Conditional Weibull pdf | 0 | 0 |
| MCMC | <u>12.77</u> | 54.29 |
| MC-Dropout | 6.27 | 15.91 |
| **MCMC + MC-Dropout** | **12.79** | <u>54.54</u> |
| XGBoost | 11.9 | 46.10 |
| SVI MFA | 11.84 | **78.86** |

Furthermore, we evaluate a time-dependent concordance index $C^{td}$ (equation (15)) [7] to see if randomly selected pairs in our data where the weapon system with the shorter observed failure time has a higher probability of failure predicted by the model. We follow Somers'd method which treats ties in time as incomparable, but pairs tied in probability of failure as .5 counts [19]:

$$C^{td} = \frac{\sum_{i \neq j} \left( \frac{1}{2} \mathbb{I}[f_i(x_i, \Delta_j) = f_j(x_j, \Delta_j)] + \mathbb{I}[f_i(x_i, \Delta_j) < f_j(x_j, \Delta_j)] \right) \mathbb{I}[\Delta_i > \Delta_j] y_j}{\sum_{i \neq j} \mathbb{I}[\Delta_i > \Delta_j] y_j}$$
$$\approx p[f_i(x_i, \Delta_j) < f_j(x_j, \Delta_j) | \Delta_i > \Delta_j], \quad (15)$$

where $\Delta$ is with respect to tslrt (and is halved if the weapon system event is a failure), $f_i(x_i, \Delta_j) = p[T < t_i + \Delta_j | T > t_i, x_i]$, and $y_j = 1$ if a failure event occurs and 0 otherwise. We evaluate the time-dependent concordance index with respect to the baseline conditional Weibull pdf (Table 2). We note that XGBoost may be lower in time dependent concordance index due to how ties in score are counted, since XGBoost does not use the time variable in many leaf splits and exhibits step-like functions for probability predictions. Furthermore, when many features were used for SVI MFA, there was convergence issues leading to worse than expected performance. This may be from the variance in computing the ELBO and consequentially the gradients from particle sampling.

Since Neural Networks are considered black-boxes with little interpretability, we implement SHaply Additive exPlanations (SHAP) [20] to understand which features of our dataset are globally important for model classification (Figure 5).

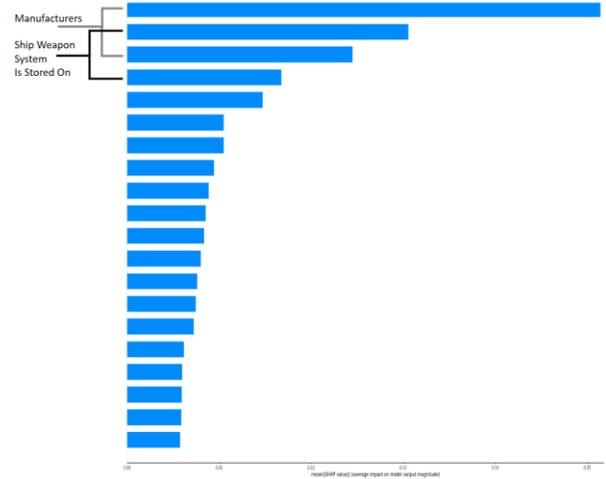

*Figure 3. SHapley Additive exPlanations (SHAP) Feature Importance Plot for MCMC + MC-Dropout Cox-Weibull Model*

The top four input features are in alignment with our Navy SMEs beliefs.

## 9 CONCLUSION

We successfully incorporated weapon system features in reliability models while ensuring the weapon system reliability model is Bayesian on the Neural Network weights and the Weibull parameters. We significantly improve weapon systems predictive maintenance metrics from the previous reliability methods, and even outperform XGBoost which is the default classifier of choice for relational databases.

## 10 FUTURE WORK

There are several directions we wish to explore in future work to improve our model design and performance. Currently, since the weapon system is repaired in maintenance/storage and not released back into the population unless it has passed recertification test, our conditional survival model assumes the weapon system did not fail up to $t_1$. To improve this assumption, correlating repeated failures for the same weapon system, we can add a multiplicative exponential term to the failure rate $\lambda$ such that after every "refresh" at certification the failure rate increases:

$$\lambda' = (\beta)^{|\{i \ : all\ failures\ for\ specified\ missile\ \cap\ t_i < t_2\}|} \lambda \quad (15)$$

which creates an accelerated failure model with $\beta > 1$. Alternatively, the introduction of an improved design to fix the failure mode can result in a lower failure rate, with $0 < \beta < 1$.

MC-Dropout [18] is known to provide a poor variational approximation for the Gaussian posterior of the Neural Network weights but is "cheap" computationally. We wish to explore stronger methods acquire distributions on Neural Network parameters such as Laplace approximations of linearized Neural Networks [21] or ensemble methods such as MultiSWAG [22].

The most interesting future work we will develop is how to model the intra/inter correlation between repeat events and/or use time-dependent covariate analysis. We will study inter correlations between failures within subpopulations of the same weapon system with different subcomponents using Bayesian hierarchical models, and intra correlations with longitudinal analysis.

## 11 SOFTWARE

All code was written in Python using python packages MLFlow, PyTorch, NumPy, Pandas, PyMC, sklearn, lifelines, scikit-survival, pyro, shap, and xgboost.

## ACKNOWLEDGEMENTS

Michael Potter and Dr. Benny Cheng were funded by Naval Innovative Science & Engineering (NISE) 6.1 basic research funding. We would like to thank the anonymous referees for their detailed comments and improvements to the paper, and to Nicole Chik and Edward Schuberg for their reviews. We additionally thank Van Nguyen for continued support and leadership.

## BIOGRAPHIES

Michael L. Potter, MS,
Department Acquisition Readiness 43
Naval Surface Warfare Center - Corona
1999 Fourth St
Norco, California 92860 USA

e-mail: michael.l.potter40.civ@us.navy.mil

Michael Potter has worked as an Electronics Engineer at the Naval Surface Warfare Center – Corona Division for over a year developing Machine Learning models. He has earned his bachelors and masters degree in Electrical and Computer Engineering at Northeastern University in 2020, and his second masters degree at the University of California Los Angeles in Electrical and Computer Engineering in 2022. Michael Potter's research interests are in recommendation systems, computer vision, linear dynamics, and deep learning.

Benny N. Cheng, PhD
Department Acquisition Readiness 43
Naval Surface Warfare Center - Corona
1999 Fourth St
Norco, California 92860 USA

e-mail: benny.n.cheng.civ@us.navy.mil

Benny Cheng is a senior Scientist at the Naval Surface Warfare Center, Corona Division, a component of the Naval Sea Systems Command, and is the US Navy's only independent analysis and assessment center. He earned his doctoral degree in Mathematics at the Massachusetts Institute of Technology in 1987, a doctoral degree in Applied Statistics at the University of California, Santa Barbara in 1995, and his bachelor's degree at the University of California, Berkeley. Prior to this position, he was a scientist at the NASA Jet Propulsion Laboratory conducting research in spectral analysis and oceanography. Dr. Cheng's current research activities are centered mainly on reliability and reliability engineering

# IEEE Copyright Notice